\documentclass{ifacconf}

\usepackage{algorithmic}
\usepackage{algorithm}
\usepackage{amsmath}
\usepackage{amssymb}
\usepackage{color}
\usepackage{lipsum}
\usepackage{graphicx}      
\usepackage{natbib}        

\newtheorem{Theorem}{\textbf{Theorem}}
\newtheorem{Lemma}{\textbf{Lemma}}

\newtheorem{Definition}{\textbf{Definition}}
\newtheorem{Remark}{\textbf{Remark}}
\newtheorem{Notation}{\textbf{Notation}}

\newcommand{\NX}{n_\mathrm{x}}
\newcommand{\NY}{n_\mathrm{y}}
\newcommand{\NU}{n_\mathrm{u}}
\newcommand{\NP}{n_\mathrm{p}}

\newcommand{\Pset}{\mathbb{P}}

\newcommand{\gfunction}[3]{g(#1,#2,#3)}

\newcommand{\red}[1]{\textcolor{black}{#1}}

\newcommand{\blue}[1]{\textcolor{black}{#1}}

\newcommand{\Norm}[1]{\|{#1}\|}
\newcommand{\Normind}[2]{\|{#1}\|_{#2}}

\newcommand{\xbar}[1]{\bar{x}(#1)}
\newcommand{\xhat}[1]{\hat{x}(#1)}
\newcommand{\Ap}[1]{A(p(#1))}
\newcommand{\AP}{A(\bfp)}
\newcommand{\APhat}{\hat{A}(\bfp)}
\newcommand{\Aphat}[1]{\hat{A}(p(#1))}
\newcommand{\Ahat}[1]{\hat{A}(#1)}
\newcommand{\Bp}[1]{B(p(#1))}
\newcommand{\Bhat}[1]{\hat{B}(#1)}
\newcommand{\Bphat}[1]{\hat{B}(p(#1))}

\newcommand{\Chat}[1]{\hat{C}(#1)}
\newcommand{\Tp}[1]{T_{\hat{\Sigma},\Sigma}(p(#1))}
\newcommand{\Tpinv}[1]{T_{\hat{\Sigma},\Sigma}(p(#1))^{-1}}
\newcommand{\Abf}{\mathbf{A}}
\newcommand{\Ahatbf}{\hat{\mathbf{A}}}
\newcommand{\Bbf}{\mathbf{B}}
\newcommand{\Bhatbf}{\hat{\mathbf{B}}}
\newcommand{\Cbf}{\mathbf{C}}
\newcommand{\Chatbf}{\hat{\mathbf{C}}}

\newcommand{\tks}{t_{s}^k}
\newcommand{\tke}{t_{e}^k}
\newcommand{\tki}{t_{i}^k}
\newcommand{\tkem}{{t^{k-1}_{e}}}

\providecommand{\abs}[1]{\lvert#1\rvert}

\newcommand{\bfp}{\mathbf{p}}

\begin{document}
\begin{frontmatter}
	
	\title{Comparing global input-output behavior of frozen-equivalent LPV state-space models} 
	\author[First,Second]{Ziad Alkhoury}
	\author[Third]{Mih\'aly Petreczky}
	\author[First]{Guillaume Merc\`{e}re}
	
	\address[First]{University of Poitiers, {L}aboratoire d'{I}nformatique et d'{A}utomatique pour les {S}yst\`{e}mes, B\^{a}timent B25, 2 rue Pierre Brousse, TSA 41105, 86073 Poitiers Cedex 9. (e-mails: ziad.alkhoury@univ-poitiers.fr, guillaume.mercere@univ-poitiers.fr)}
	\address[Second]{Ecole des Mines de Douai, F-59500 Douai, France.}
	\address[Third]{CNRS, Centrale Lille, UMR 9189 - CRIStAL-Centre de Recherche en Informatique, Signal et Automatique de Lille, F-59000 Lille, France. (e-mail: mihaly.petreczky@ec-lille.fr)}
	\thanks{This work was partially supported by ESTIREZ project of Region Nord-Pas de Calais, France.}
	

	\begin{abstract}
         It is known that in general, \emph{frozen equivalent}  (Linear Parameter-Varying) LPV models, \emph{i.e.},  LPV models which have the same input-output behavior for each constant scheduling signal, might exhibit different input-output behavior for non-constant scheduling signals. In this paper, we provide an analytic error bound on the difference between the input-output behaviors of two LPV models which are frozen equivalent.  This error bound turns out to be a function of both the speed of the change of the scheduling signal and the  discrepancy between the coherent bases of the two LPV models. In particular, the difference between the outputs of the two  models can be made arbitrarily small by choosing a scheduling signal which changes slowly enough. An illustrative example is presented to show that the choice of the scheduling signal can reduce the difference between  the input-output behaviors of frozen-equivalent LPV models.
  %

	\end{abstract}
\end{frontmatter}
	\section{INTRODUCTION}
	
	As shown, \emph{e.g.}, in \cite{Toth2010, Lal11}, existing techniques dedicated to the identification of Linear Parameter-Varying (LPV) systems can be split up into two main families: the local and the global approach. On the one hand, the global approach focuses on a global procedure for which it is assumed that one global experiment can be performed in which the control inputs as well as the scheduling variables can be both excited (see, \emph{e.g.}, \cite{Bamieh2002,Felici2007}). By construction, these techniques are restricted to specific systems where the components of the scheduling variables are totally controllable and excitable (and not only measurable). On the other hand, the local approach is based on a multi-step procedure (see, \emph{e.g.},
	 \cite{Vizer2014a},  \cite{Val13}, \cite{LM07}) where first, frozen linear time-invariant models are estimated for constant values of the the scheduling variables, then the global LPV model is built from the interpolation of the local LTI models. \blue{ A substantial amount of local techniques are available in the literature (see \cite{DeCaigny2011}, \cite{Luspay2011}). This is probably due to the fact that the identification techniques for LTI systems are more mature than the LPV ones, in addition to the development of algorithms to optimize the number of local operating points involved in the procedure \cite{Vizer2014a}, as well as their strong link with the gain scheduling procedure \cite{Rugh2000}. However, it is clear that none of these techniques is consistent when black-box state-space LPV models are estimated.} As pointed out first in \cite{Tal07b}, the main reasons why the current solutions fail to mimic the global behavior of the LPV system are due to $(i)$ the lack of information about the dynamical behavior of the scheduling variables $p$ in the frozen models when  LPV models depending on $p$ dynamically are handled, $(ii)$ the difficulty to express the local LTI models w.r.t. a coherent basis \cite{Toth2010} before the interpolation. While solutions have been suggested to bypass these difficulties when gray-box state-space LPV models are considered \cite{Val13, Val13b, VML15},  restricting slow variations of $p$ (w.r.t. the dynamics of the system) is the only assumption to make  in order to guarantee a reliable global model behavior when black-box models are handled, \emph{i.e.}, when the only source of information is restricted to measured data, see \cite{Kal09}.

In this paper, we present an analytical error bound on the difference between the global input-output behavior of
two frozen equivalent LPV models. By global input-output behavior we mean the input-output behavior along \emph{all} scheduling signals, while by frozen equivalence of two LPV models we mean that their input-output behavior is the same 
along all \emph{constant} scheduling signals. 
As it was discussed above, even if the data were generated by an LPV model, 
local methods for system identification in general result in LPV models which 
are only frozen equivalent to the `true' one. Hence, the error bound  of this paper allow us 
\begin{enumerate}
\item
to evaluate the difference between the
input-output behavior LPV model obtained by local identification methods, and the `true' LPV model which generated the data,

\item
 to evaluate the difference between the input-output behavior of LPV models obtained by different local identification methods.
\end{enumerate}
The error bound includes the speed of the change of the scheduling signal, the minimal time the scheduling signal remains constant (can be $1$ time step in general) and a measure of inconsistency of the bases of the frozen LTI models. Note that this bound is
valid only for (quadratically) stable LPV models. 
It turns out that even if the frozen LTI models are not in a coherent basis, the error between the input-output behaviors of the
corresponding LPV models can be kept arbitrary small by using sufficiently slow scheduling signals, or by keeping the scheduling
signal constant for long enough. 
%
%
 This allows us to conclude that the problem of choosing the `correct' state transformation required for the interpolation step becomes less critical for slowly varying scheduling signals. 

While the results of the paper seem to be intuitive, they have never been presented formally, to the best of our knowledge. A precise error bound could be  of great interest for system identification and control design for LPV models, because it 
allows the user to quantify the effect of the modeling error due to local identification on the system performance. In turn, this
could be incorporated as an additional constraint in control synthesis, or used as a minimization criteria in parametric system identification.

	 
	 
	 \textbf{Outline of the paper.} In Section~\ref{sec:problemFormulation}, we present the basic notions to set up the framework of the paper. Next, in Section~\ref{sec:mainResults}, we present the main results.
 Section~\ref{sec:illustrativeExample} presents an illustrative example. Finally, Section~\ref{sec:conclusion} concludes the paper. 
The proofs of the results are presented in Appendix.

	\section{Problem formulation}\label{sec:problemFormulation}
	To set up the framework of the paper, we give some basic definitions for LTI and LPV models.
	\begin{Definition}[LTI models]
		A discrete-time Linear Time-Invariant (LTI) model  is defined as follows
			\begin{equation}
			\label{equ:LTISystem}
			\mathcal{L}\left\{
			\begin{array}{lcl}
			x(t+1) &=&  \mathbf{A} x(t) + \mathbf{B} u(t) , \\
			y(t) &=& \mathbf{C}x(t).
			\end{array}
			\right. 
			\end{equation}
			 $x(t) \in \mathbb{X}=\mathbb{R}^{n_\mathrm{x}}$ is the state at time $t$, $y(t) \in \mathbb{Y}=\mathbb{R}^{n_\mathrm{y}}$ is the output at time $t$, $u (t) \in \mathbb{U} = \mathbb{R}^{n_\mathrm{u}}$ is the input at time $t$, 
and $\mathbf{A} \in \mathbb{R}^{n_{\mathrm x}\times n_{\mathrm x}}, \mathbf{B}  \in \mathbb{R}^{n_{\mathrm x} \times n_{\mathrm u}}$ and $\mathbf{C}  \in  \mathbb{R}^{n_{\mathrm y} \times n_{\mathrm x}}$ 
			In the sequel, we will use the notation
			\begin{equation}
			\mathcal{L}=(\mathbf{A}, \mathbf{B}, \mathbf{C})
			\end{equation}
			for an LTI model of the form \eqref{equ:LTISystem}.
	\begin{Definition}[LPV models]
	A discrete-time Linear Parameter Varying (LPV) model $(\Sigma)$ is defined as follows
	\begin{equation}
	\label{equ:LPVSystem}
	\Sigma\left\{
	\begin{array}{lcl}
	x(t+1) &=&  A(p(t)) x(t) + B(p(t)) u(t) , \\
	y(t) &=& C(p(t)) x(t).
	\end{array}
	\right. 
	\end{equation}
\end{Definition}
	$x(t) \in \mathbb{X}=\mathbb{R}^{n_\mathrm{x}}$ is the state at time $t$, $y(t) \in \mathbb{Y}=\mathbb{R}^{n_\mathrm{y}}$ is the output at time $t$, $u (t) \in \mathbb{U} = \mathbb{R}^{n_\mathrm{u}}$ is the input  at $t$
	 and $p(t) \in \mathbb{P}$ is the value of the scheduling signal at time $t$. 
	The matrix functions $A : \Pset \mapsto \mathbb{R}^{n_{\mathrm x}\times n_{\mathrm x}}, B : \Pset \mapsto  \mathbb{R}^{n_{\mathrm x} \times n_{\mathrm u}}$ and $C : \Pset \mapsto  \mathbb{R}^{n_{\mathrm y} \times n_{\mathrm x}}$  
	are assumed to be continuous.  We assume that $\mathbb{P}$ is compact, \emph{i.e.}, closed and bounded set.	In the sequel, we will use the short notation
	\begin{equation}
	\Sigma=(A(.), B(.), C(.))
	\end{equation}
	to denote a model of the form \eqref{equ:LPVSystem}. Notice that we use the notation $A(.), B(.)$ and $C(.)$ to emphasize that these matrices are maps, not constant matrices.
		\end{Definition}
	
	\begin{Remark}
		In order to avoid notational ambiguity, we denote that matrices of the LTI models by bold symbols, \emph{i.e.}, $(\Abf, \Bbf, \Cbf)$, while we denote that matrices of the LPV model by $(A(.), B(.), C(.))$.
	\end{Remark}
	
	From now on, we will use the following notation:
	for a set $\mathbb{A}$, we denote by $\mathbb{A}^{\mathbb{N}}$ the set of all functions of the form $\phi:\mathbb{N} \rightarrow \mathbb{A}$. An element of $\mathbb{A}^{\mathbb{N}}$ can be thought of as a signal in discrete-time. 
	By a solution  of $\mathcal{L}$ (respectively $\Sigma$), we mean a tuple of trajectories $(x,y,u)\in(\mathcal{X},\mathcal{Y},\mathcal{U})$ satisfying \eqref{equ:LTISystem},  (respectively $(x,y,u,p)\in(\mathcal{X},\mathcal{Y},\mathcal{U},\mathcal{P})$ satisfying \eqref{equ:LPVSystem}), for all $k \in \mathbb{N}$, where 
	$\mathcal{X}=\mathbb{X}^\mathbb{N}, \mathcal{Y}=\mathbb{Y}^{\mathbb{N}}, \mathcal{U}=\mathbb{U}^{\mathbb{N}},\mathcal{P}=\mathbb{P}^\mathbb{N}$.	
		Note that $\mathcal{P}$ denotes the set of scheduling signals, while $\mathbb{P}$ denotes the set of all possible values of the scheduling parameters. That is,  an element of $\mathcal{P}$ is a sequence, while an element of $\mathbb{P}$ is a vector. The same remark holds for $\mathcal{X},\mathcal{Y},\mathcal{U}$ and $\mathbb{X},\mathbb{Y},\mathbb{U}$ respectively. Note that in the sequel, unless stated otherwise, \emph{we will use $p$ to denote a scheduling signal, and we use $\bfp $ to denote a potential value of the scheduling variable, \emph{i.e.}, $p \in \mathcal{P}$ is a sequence, but $\bfp \in \mathbb{P}$ is a vector of constant values.} 
		
		 	\begin{Definition}[Input-output map of LPV models]
		 	\hspace{1pt}\\Define the functions 
		 	\begin{align}
		 	Y_{\Sigma} & : \mathcal{U} \times \mathcal{P} \mapsto \mathcal{Y},
		 	\end{align}	
		 	for an LPV model $\Sigma$, such that for any $(x,y,u,p) \in (\mathcal{X} \times \mathcal{Y}  \times\mathcal{U}  \times\mathcal{P})$, 
		 	$y=Y_{\Sigma}(u,p)$ hold if and only if $(x,y,u,p)$ satisfy \eqref{equ:LPVSystem}, and $x(0)=0$. The function 
		 	$Y_{\Sigma}$ is called the input-output map of $\Sigma$.
		 \end{Definition}

	\begin{Definition}[Frozen model of $\Sigma$ at $\bfp$]
		A frozen model of an LPV $\Sigma=(A(.), B(.), C(.))$  at $\bfp$, denoted by $\mathcal{L}_{\Sigma}(\bfp)$, is the LTI model defined by
		\begin{equation}
		\mathcal{L}_{\Sigma}(\bfp)=(A(\bfp), B(\bfp), C(\bfp)).
		\end{equation}
		
	\end{Definition}

	\begin{Definition}[Frozen-minimal LPV models]
		\hspace{1pt}\\
		An LPV model $\Sigma=(A(.), B(.), C(.))$  is frozen-minimal, if $\forall \bfp \in \mathbb{P}$, the LTI $\mathcal{L}_{\Sigma}(\bfp)=(A(\bfp), B(\bfp), C(\bfp))$ is minimal\footnote{According to the classical definition, \emph{i.e.}, $(A(\bfp), B(\bfp))$ is controllable and $(A(\bfp), C(\bfp))$ is observable, (see \cite{Rugh1996}).}.
	\end{Definition}


	 In this paper, all models have the same state, input, output and scheduling signal dimensions. Therefore, the matrices and vectors dimensions $\NX, \NP, \NY$ and $\NU$ will be omitted when clear from the context. 

	\begin{Definition}[Frozen-equivalent LPV models at $\bfp$]
		\hspace{1pt}\\ We say that two LPV models $\Sigma=(A(.),B(.),C(.))$ and $\hat{\Sigma}=(\Ahat{.},\Bhat{.},\Chat{.})$ are frozen equivalent, if $\forall \bfp \in \Pset$ the LTI models $\mathcal{L}_{\Sigma}(\bfp)$ and $\mathcal{L}_{\hat{\Sigma}}(\bfp)$ have
 the same transfer function, \emph{i.e.}, for all $s \in \mathbb{C}$ which is not an eigenvalue of $A(\bfp)$ or $\Ahat{\bfp}$,
$C(\bfp)(sI_{\NX}-A(\bfp))^{-1}B(\bfp) = \Chat{\bfp}(sI_{\NX}-\Ahat{\bfp})^{-1}\Bhat{\bfp}$. 
	\end{Definition}
	That is, if $\Sigma$ and $\hat{\Sigma}$ are frozen input-output equivalent  LPV models, then, for any $p \in \mathcal{P}$ such that $p$ is constant (\emph{i.e.}, $p(0) = p(k), \forall k \in \mathbb{N}$), and for any $u \in \mathcal{U}$,
		$Y_{\Sigma}(u, p) =Y_{\hat{\Sigma}}(u,p)$.
	
	
	
%

\section{Main results}\label{sec:mainResults}
	In this section, the difference between the outputs of two frozen-minimal and frozen-equivalent LPV models 
is characterized. 
	
	\subsection{Preliminaries}\label{sec:preliminaries}
	
	In order to compare the global behavior of the LPV models $\Sigma$ and $\hat{\Sigma}$, we need to introduce  some concepts. We start from the fact that there is a unique isomorphism between any two input-output equivalent minimal LTI models (\emph{i.e.}, they have the same transfer function) $\mathcal{L}$ and $\hat{\mathcal{L}}$ (see \cite{Rugh1996}). This means that there is a unique nonsingular $\NX \times \NX$ matrix $T$ that fulfills
	\begin{align}
	T \Abf T^{-1} = \Ahatbf, & & T\Bbf = \Bhatbf, & & \Cbf T^{-1} = \Chatbf.
	\end{align}
	where $\{\Abf, \Bbf, \Cbf\}$ and $\{\Ahatbf, \Bhatbf, \Chatbf\}$ are the state-space matrices of  $\mathcal{L}$ and $\hat{\mathcal{L}}$ respectively. We say that $T$ is an isomorphism from $\mathcal{L}$ to $\hat{\mathcal{L}}$.

	Now, let us assume that  $\mathcal{L}_{\Sigma}(\bfp)=({A}(\bfp), {B}(\bfp), {C}(\bfp))$  and $\mathcal{L}_{\hat{\Sigma}}(\bfp)=(\hat{A}(\bfp), \hat{B}(\bfp), \hat{C}(\bfp))$ are two frozen LTI models of $\Sigma$ and $\hat{\Sigma}$, respectively, at $\bfp \in \Pset$.
	We assume that $\Sigma$  and $\hat{\Sigma}$ are both frozen-equivalent and frozen-minimal LPV models, then $\mathcal{L}_{\Sigma}(\bfp)$ and $\mathcal{L}_{\hat{\Sigma}}(\bfp)$ are input-output equivalent and both are minimal for all $\bfp \in \Pset$. Hence, there exists a (unique) isomorphism $T(\bfp)$ from $\mathcal{L}_{\hat{\Sigma}}(\bfp)$ to $\mathcal{L}_{\Sigma}(\bfp)$ for all $\bfp \in \Pset$. 
%
	Let us call the map $T_{\hat{\Sigma}, \Sigma}: \bfp \rightarrow T(\bfp)$, the \emph{frozen isomorphism} map from $\hat{\Sigma}$ to $\Sigma$.
	\begin{Lemma}\label{lem:continuousIsomorphism}
		Let $\Sigma, \hat{\Sigma}$  be two frozen-minimal and frozen-equivalent LPV models
		.
		 If $A, B, C,\hat{A},\hat{B}$ and $\hat{C}$ are continuous maps, then $T_{\hat{\Sigma}, \Sigma}$ is a continuous map as well.
	\end{Lemma}
	The proof of Lemma~\ref{lem:continuousIsomorphism} can be found in Appendix~\ref{prf:continuousIsomorphism}
	
	\begin{Remark}\label{rem:ALPVisomorphism}
	It is important to note that $T_{\hat{\Sigma},\Sigma}$ is not an LPV isomorphism between the two LPV models. In order for 
        $T_{\hat{\Sigma},\Sigma}$ to be an LPV isomorphism, 
		\(\Ap{t} = T_{\hat{\Sigma},\Sigma}(p(t+1))\Aphat{t}T_{\hat{\Sigma},\Sigma}^{-1}(p(t)),\)
		should hold for any scheduling signal $p \in \mathcal{P}$, see \cite{Kulcsar2011,Toth2010} for more details.
	\end{Remark}
	In the rest of the paper, we will need the following notations.
	\begin{Notation}
		In the sequel, $\Norm{A}$ will denote the induced 2-Norm of any matrix $A$, while for a vector $\bfp$,  $\Norm{\bfp}$ will denote the Euclidean vector norm of $\bfp$,  (see \cite{Meyer2000}). We denote by $I_{\NX}$ the $\NX \times \NX$ identity matrix.
	\end{Notation}
	To proceed with the characterization of the global behavior equivalence, let us define
	\begin{align}
	\forall \bfp_1,\bfp_2 \in \Pset:  M_{\bfp_1,\bfp_2} = T_{\hat{\Sigma},\Sigma}(\bfp_1)T_{\hat{\Sigma},\Sigma}(\bfp_2)^{-1}.
	\end{align}
	Notice that, if $T(\bfp_1)=T(\bfp_2)$, then, $M_{\bfp_1,\bfp_2} = I_{\NX}$. Define the coefficients $K_B, K_C$ and $K_T$ as follows,
	\begin{align}
	\sup\limits_{\bfp \in \Pset} \Norm{B(\bfp)} &= K_B, \label{eq:def_kb} \\
	\sup\limits_{\bfp \in \Pset} \Norm{C(\bfp)} &= K_C,\label{eq:def_kc}\\
	\sup\limits_{\bfp \in \Pset} \Norm{T_{\hat{\Sigma},\Sigma}(\bfp)} &=   K_T. \label{eq:def_kt}
	\end{align}
	It is important to note that $K_B, K_C$ and $K_T$ are finite. This follows directly from the fact that each of the maps $B, C$ and $T_{\hat{\Sigma},\Sigma}$ are continuous, and $\Pset$ is a compact set.
	
	Now, for every scheduling signal $p \in \mathcal{P}$, define
	\begin{align}
	 K_M(p) = \sup\limits_{k \in \mathbb{N}}  \Norm{I_{\NX}-M_{p(k),p(k+1)}}.\label{eq:km}
	\end{align}
	Then, 
	\begin{align}
		K_M(p) \leq \sup\limits_{\bfp_1,\bfp_2 \in \mathbb{P}} \Norm{I_{\NX}-M_{\bfp_1,\bfp_2}} < +\infty .
	\end{align}
	Notice that $\sup\limits_{\bfp_1,\bfp_2 \in \mathbb{P}} \Norm{I_{\NX}-M_{\bfp_1,\bfp_2}}$ exists because $T_{\hat{\Sigma},\Sigma}$ is continuous, and hence $\Pset \times \Pset \ni (p_1,p_2) \mapsto M_{p_1,p_2}$ is continuous and defined on a compact set. This means that, $\exists K_M > 0$, such that
	\begin{align}
		\forall p \in \mathcal{P}: K_M(p) \leq K_M.
	\end{align} 

 	Let us also assume that $\Sigma$ and $\hat{\Sigma}$, are quadratically stable, \emph{i.e.}, there exists a positive definite matrix $P > 0$, such that
        for all $\bfp \in \Pset$, $\AP^T P \AP - P < 0$ and for $\hat{A}(\bfp)$ similarly.
       \begin{Lemma}\label{lem:quadraticStability}
		If $\Sigma$ is quadratically stable, then there exists a nonsingular $S \in \mathbb{R}^{\NX \times \NX}$, such that
                the LPV
		\begin{align*}
			\bar{\Sigma} = (S A(.) S^{-1}, S B(.), C(.)S^{-1})
		\end{align*}
                 satisfies
		\begin{align*}
		\sup\limits_{\bfp \in \mathbb{P}}\Norm{S A(\bfp) S^{-1}} < \alpha < 1,
		\end{align*}
                for some $\alpha \in (0,1)$. 
	\end{Lemma}
	In the sequel, we will assume that
	\begin{align}
	\sup\limits_{\bfp \in \Pset} \Norm{\AP} \leq \alpha < 1,\label{eq:alphaA} \\ \sup\limits_{\bfp \in \Pset} \Norm{\Ahat{\bfp}} \leq \hat{\alpha} < 1. \label{eq:alphaHatA}
	\end{align}
      If it is not the case, then we can replace  $\Sigma$ by $\bar{\Sigma}$. 
     Note that $\alpha$ in  \eqref{eq:alphaA} is an upper bound on the Lyapunov exponent of  dynamical system $x(t+1)=A(p(t))x(t)$; the smaller $\alpha$ is, the faster the solutions of $x(t+1)=A(p(t))x(t)$ converge to zero. 

	If $\hat{\Sigma}$ satisfies  \eqref{eq:alphaA}, then 
        it follows from \cite[Proof of Lemma 27.4 and Theorem 27.2]{Rugh1996} and 
        \eqref{eq:def_kb} -- \eqref{eq:def_kc}
	that there exists $\mu_1 > 0$ such that
       	for any solution $(\hat{x},\hat{y},u,p)$ of $\hat{\Sigma}$ such that $x(0)=0$ and $\sup_{t \in \mathbb{N}} \|u(t)\| < +\infty$, 
	\begin{align}\label{eq:def_mu1}
		\sup\limits_{t} \Norm{\xhat{t}} \leq& \,\, \mu_1 \sup_{t \in \mathbb{N}} \|u(t)\|. 
        \end{align}

	From now on, when comparing the outputs of two LPV models, we will assume that they are subject to the same bounded input from the set
	
		\begin{align*}
			l_{\infty}(\mathbb{U}) = \{u \in  \mathbb{U}: \Normind{u}{l_{\infty}}= \sup_{t \in \mathbb{N}} \|u(t)\| < +\infty\}.
		\end{align*}
 Moreover, we will consider piecewise-constant scheduling signals from the following set. 
	\begin{align}\label{eq:sDelta}
	\mathcal{S}_{\Delta} \!\!=\!\! \{p\in \Pset\!:\! \forall k \! \in  \mathbb{N},\! \forall i \in \{0,1, \cdots,\Delta-1\}\!:\! & \\ & \hspace{-100pt}\nonumber
	p(k \Delta + i) \!\!=\!\! p(k \Delta)\!\},
	\end{align}\normalsize
	
\subsection{General error bound}\label{subsec:differencecharacterization}
We are now ready to present our main result on the global input-output equivalence of frozen-equivalent LPV models.
	\begin{Theorem}[Difference characterization]\label{th:differenceCharacterization}
		Let $\Sigma$ and $\hat{\Sigma}$ be two frozen-minimal, frozen-equivalent and quadratically stable  LPV models.
		Then, for all $p \in \mathcal{S}_{\Delta}, \Delta > 0$ and 
                $u \in l_{\infty}(\mathbb{U})$, $t \in \mathbb{N}$, 
		\begin{align}\label{eq:outputEquation}
		\Norm{Y_{\Sigma}(u, p)(t)-Y_{\hat{\Sigma}}(u,p)(t)} & \nonumber \\ 
                 &  \hspace{-70pt} \leq \gfunction{\Delta}{K_M(p)}{t~ \mathrm{mod}~ \Delta } \cdot \Normind{u}{l_{\infty}}\nonumber\\ & \hspace{-70pt} \leq \gfunction{\Delta}{K_M}{t~ \mathrm{mod}~ \Delta}. \Normind{u}{l_{\infty}}.
		\end{align}

	where $t~ \mathrm{mod}~ \Delta$ denotes the remainder of dividing $t$ by $\Delta$, and for all $i=0,\ldots,\Delta-1$,
		\begin{align}\label{eq:gDefinition}
		\gfunction{\Delta}{K}{i}  = \frac{\alpha^{i}}{1-\alpha^{\Delta}}K (\alpha K_T \mu_1 + K_B) K_C,
		\end{align}
		and $K_B, K_C, K_T, K_M(p),\alpha$ and $\mu_1$ are defined in equations \eqref{eq:def_kb}, \eqref{eq:def_kc}, \eqref{eq:def_kt}, \eqref{eq:km}, \eqref{eq:alphaA} and \eqref{eq:def_mu1} respectively.
	\end{Theorem}
	The proof of Theorem~\ref{th:differenceCharacterization} can be found in Appendix~\ref{prf:th:differenceCharacterization}.
		 Eq.~\eqref{eq:outputEquation} gives a conservative bound on the output error of two LPV systems whose frozen transfer
functions are the same.  
It clearly shows that the error is affected by each of the following:
			\begin{itemize}
				\item The length of the interval $\Delta$, on which the  scheduling signal $p$ is constant.
				\item The Lyapunov exponent $\alpha$ of the system, which expresses the degree of stability.
				\item The distance $K_M(p)$ which expresses  the degree of inconsistency of the basis of the frozen
                                      models: if all the frozen models are in a consistent basis, then $K_M(p)$ is zero.
				\item The amplitude of the input $\Normind{u}{l_{\infty}}$. Note that \eqref{eq:outputEquation} remains true if instead of $\Normind{u}{l_{\infty}}$ we take $\Normind{u}{l_{2}}=(\sum_{k=0} \|u(k)\|^2_2)^{\frac{1}{2}}$. 	
			\end{itemize}
	\subsection{Special cases}\label{sec:detailedResults}
	In this subsection, we formally derive and present important results based on Theorem~\ref{th:differenceCharacterization}, \emph{i.e.}, which aim at showing the effects  of different variables on the difference of the input-output map presented in Equations~\eqref{eq:outputEquation} and \eqref{eq:gDefinition} from Subsection~\ref{subsec:differencecharacterization}. 
	\begin{Theorem}[Switching interval]\label{th:switchingInterval}
		Let $\Sigma$ and $\hat{\Sigma}$ be two frozen-minimal, frozen-equivalent and quadratically stable  LPV models. Then,  for all
   $\varepsilon > 0$ there exists $\Delta_m > 0$ such that for all $p \in \mathcal{S}_{\Delta}$, $\Delta > \Delta_m$, 
       $u \in l_{\infty}(\mathbb{U})$, and for all $t \in \mathbb{N}$ with $t~ \mathrm{mod}~ \Delta > \Delta_m$,
		\begin{align}
		 \Norm{Y_{\Sigma}(u, p)(t) -Y_{\hat{\Sigma}}(u,p)(t)}< \varepsilon . \Normind{u}{l_{\infty}}.
		\end{align}
	\end{Theorem}
	The proof of Theorem~\ref{th:switchingInterval} can be found in Appendix~\ref{prf:th:switchingInterval}.

	
	Theorem~\ref{th:switchingInterval} tells us that if we have a piecewise constant scheduling signal, and the duration of each constant piece is large enough, then on each constant piece the outputs can get arbitrarily close, no matter how bad we choose the coordinates of the frozen models. In case of switching sequences we call the length of the interval on which $p$ is constant the dwell time. For switched systems, the theorem says that for large enough dwell times, the outputs of the system can get arbitrary close towards the end of the active period of each discrete mode.
		
	
	\begin{Theorem}[Speed of change of scheduling parameter]\label{th:speedOfChange}
	 Let $\Sigma$ and $\hat{\Sigma}$ be two frozen-minimal, frozen-equivalent and quadratically stable  LPV models.
		Then, for every $\varepsilon > 0$, there exists  $\delta_m>0$ such that, for all $p \in \mathcal{P}$ satisfying
		\begin{align}
		\forall t \geq 0: &\abs{p(t+1)-p(t)} < \delta_m,
		\end{align}
		and for all $u \in l_{\infty}(\mathbb{U})$, $t \in \mathbb{N}$,
		\begin{align}
 \Norm{Y_{\Sigma}(u, p)(t) -Y_{\hat{\Sigma}}(u,p)(t)}  < \varepsilon . \Normind{u}{l_{\infty}}. \nonumber
		\end{align}
	\end{Theorem}
	The proof of Theorem~\ref{th:speedOfChange} can be found in Appendix~\ref{prf:th:speedOfChange}.
	
	 This result is particularly interesting, because the scheduling signal is no longer restricted to be a piece-wise constant signal. This implies that we can still guarantee the input-output `approximate' equivalence using a slowly changing scheduling signal, and the result holds for any $t$. 	

	However, if no conditions can be imposed on the scheduling signal, it is of great interest to evaluate the effect of the stability of the system on the difference between input-output maps of the two frozen-equivalent models. In the next theorem, we study the mentioned effect.

	\begin{Theorem}[Stability]\label{th:stability} Let $\Sigma$ and $\hat{\Sigma}$ be two frozen-minimal, frozen-equivalent and quadratically stable  LPV models. Then, for all $\varepsilon > 0$, there exists $\alpha_m>0$ such that, if $\Sigma$ satisfies
	\begin{align}\label{eq:S1P}
		\sup\limits_{\bfp \in \Pset} \Norm{\AP} \leq  \alpha_m,
	\end{align}
	then, for all $u \in l_{\infty}(\mathbb{U})$, $p \in \mathcal{P}$, $t \in \mathbb{N}$, 
	\begin{align}
	 \Norm{Y_{\Sigma}(u, p)(t) -Y_{\hat{\Sigma}}(u,p)(t)} < \varepsilon . \Normind{u}{l_{\infty}}.
	\end{align}
	\end{Theorem}
		The proof of Theorem~\ref{th:stability} can be found in Appendix~\ref{prf:th:stablility}.
		

\subsection{Discussion: basic trade-offs}\label{sec:discussion}
Notice that, from Eq.~\eqref{eq:outputEquation}, we have three factors that play important role in the error characterization.
\begin{enumerate}
	\item The switching interval of the scheduling signal. By choosing a relatively long switching interval $\Delta$, we guarantee that the difference between the outputs of the LPV models becomes small enough towards the end of each constant interval.
	\item The rapidity of change of the scheduling signal, \emph{i.e.}, the difference between two consecutive values of the scheduling variable, which is $\abs{p(t+1)-p(t)}$. This case is of great interest when $\Delta = 1$, because the scheduling variable will not be restricted to a piece-wise constant signal any longer, and it will be allowed to change at each time instant. By slowly changing the scheduling signal, we can get relatively very small difference between the input-output behaviors of frozen-equivalent LPV models.
	\item The stability of the system. The more the system is stable, the less it responds to the applied input, \emph{i.e.}, the less the difference between two LPV models is. 
\end{enumerate}
\red{The results above actually show the following. On the one hand, even if the identified LTI models are in a non-coherent state-space basis, the obtained state-space LPV model can be still an accurate approximation of the real system provided that the scheduling signal is slow enough. More precisely, using slow changing scheduling signal, or using a piece-wise constant scheduling signal with long switching intervals is enough to obtain satisfying results. On the other hand, if the basis of the frozen state-space representation is close to a consistent one, then, even for `faster' scheduling signal the approximation error will be small.}

\section{Illustrative example}\label{sec:illustrativeExample}

Let us consider the LPV model $\Sigma$ of the form \eqref{equ:LPVSystem}, such that $\NX = 2, \NP = 1, \NU = \NY = 1$, and such that, for any $\bfp \in \Pset := [0.1, 0.4]$,
\[
A(\mathbf{p})\!\!= \!\!\begin{bmatrix}0 &  0.2\mathbf{p}\\0.2  & \mathbf{p}
\end{bmatrix}, ~
B(\mathbf{p})\!\!=\!\!\begin{bmatrix}	1\\1\end{bmatrix}, ~
C(\mathbf{p})\!\!=\!\!\begin{bmatrix}1 \\ 1\end{bmatrix}^\top \!\!.\]\normalsize

In order to obtain an LPV model $\hat{\Sigma}$, which is frozen-equivalent to $\Sigma$, we perform an LPV identification process which includes the following steps:
\begin{enumerate}
	\item Input-output data generation at each possible constant value of the scheduling signal.
	\item Black-box LTI system identification using a subspace method. This step results in multiple LTI models, all are obtained in the same canonical form.
	\item Linear interpolation of the matrices of the frozen models with respect to the values of the scheduling signal. This step gives us the matrices of the LPV model $\hat{\Sigma}$, which is frozen equivalent to the original one.
\end{enumerate}

For comparing the outputs of the original LPV model and the identified one, we excite both LPV models $\Sigma$ and $\hat{\Sigma}$ using the same input $u(t)=1$ and the same scheduling signal $p$, which is different for each of the following figures. 

\begin{center}
	\begin{figure}
		\centering
		\includegraphics[trim=110 260 100 255,clip,width=0.4\textwidth]{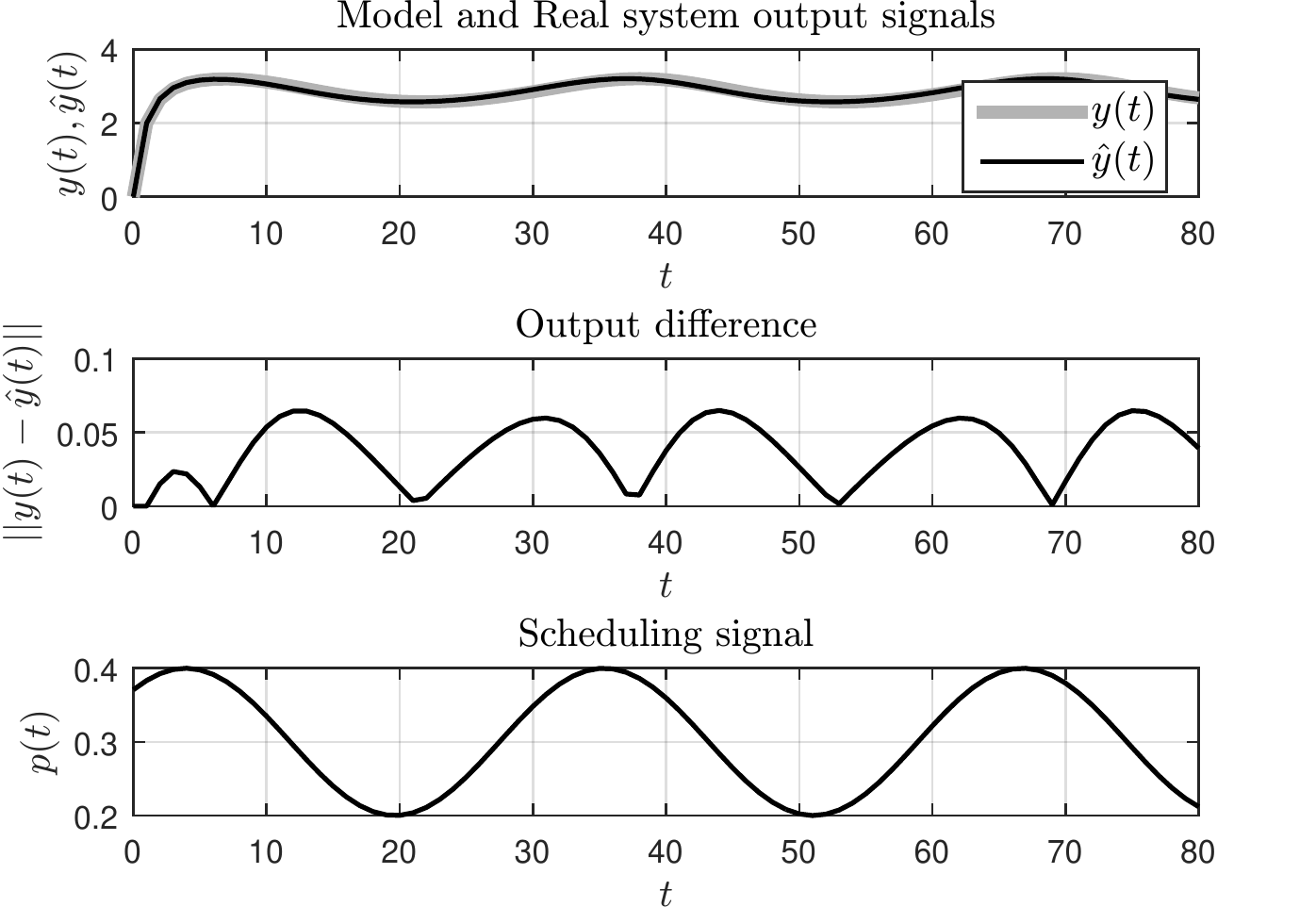}
		\caption{\small Output difference for $p(t)=0.3+0.1 sin(t/5)$. \normalsize}
		\label{fig:ex:pSin1}
	\end{figure}
\end{center}

\begin{center}
	\begin{figure}
		\centering
		\includegraphics[trim=110 260 100 255,clip,width=0.4\textwidth]{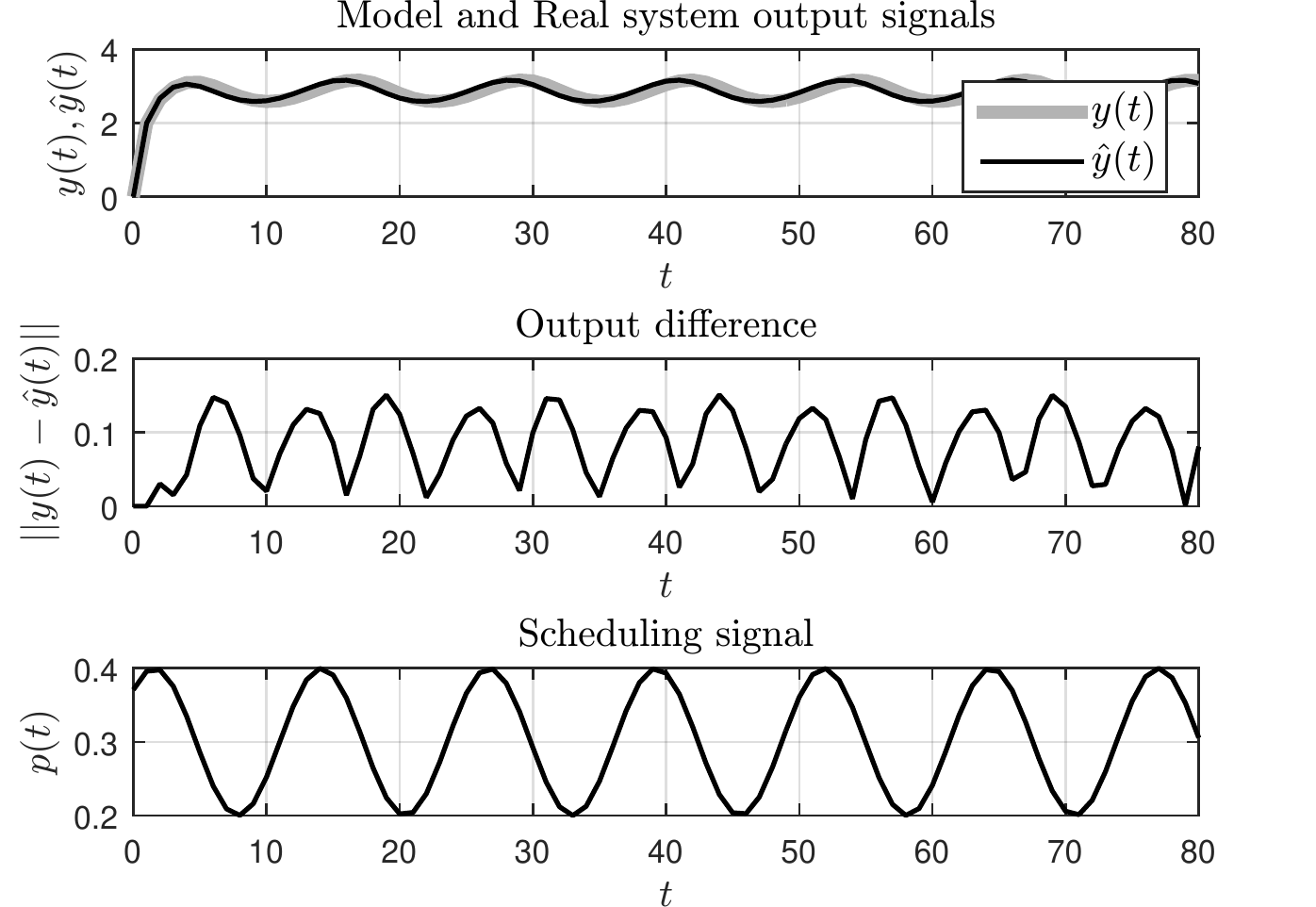}
		\caption{\small Output difference for $p(t)=0.3+0.1 sin(t/2)$. \normalsize}
		\label{fig:ex:pSin2}
	\end{figure}
\end{center}

In Fig~\ref{fig:ex:pPWC}, the scheduling signal is chosen to be a piece-wise constant one as is clearly shown in the figure. Notice that the difference between the outputs is non-zero at the switching instances of $p$, \emph{i.e.}, at $t=11, 21,..., 71$. Notice also that the difference decreases approaching zero as we go away from the switching instances. This result is expected based on Theorem~\ref{th:switchingInterval}. We can also see that the difference between the outputs depends on the jumps in the switching signal. 
\begin{center}
	\begin{figure}
		\centering
		\includegraphics[trim=110 260 100 255,clip,width=0.4\textwidth]{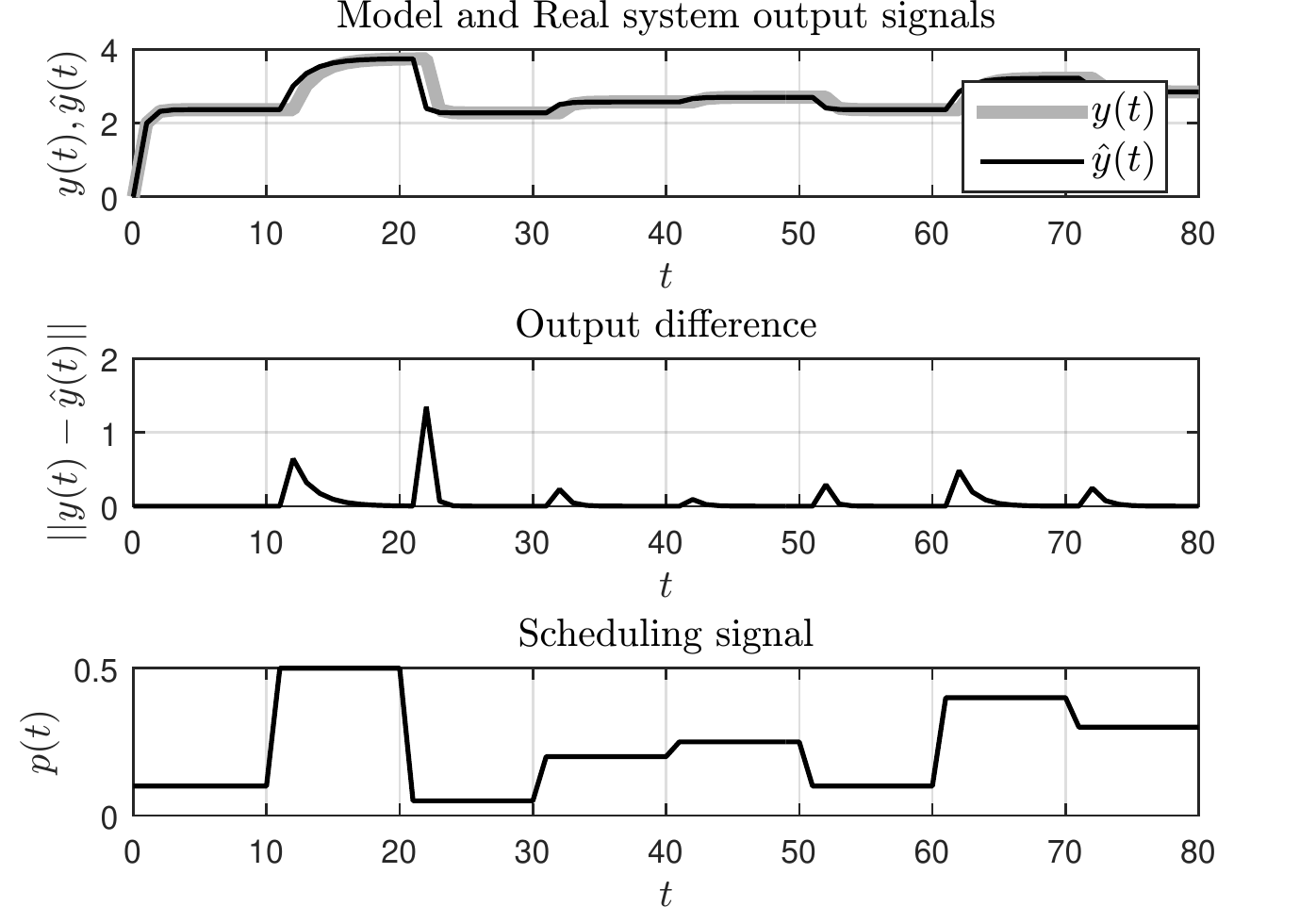}
		\caption{\small Output difference for $p$ a piece-wise constant signal. \normalsize}
		\label{fig:ex:pPWC}
	\end{figure}
\end{center}
In Fig.~\ref{fig:ex:pSin1} and \ref{fig:ex:pSin2}, the scheduling signal is chosen to be a sinusoidal one. This allows us to illustrate the results of Theorem~\ref{th:speedOfChange}. It is clear that the output difference of $\Sigma$ and $\hat{\Sigma}$ becomes bigger as we increase the frequency of the scheduling signal in Fig~\ref{fig:ex:pSin2}, in comparison to Fig~\ref{fig:ex:pSin1}.
It is important to mention that the interpolation step is a crucial one, because the resulting LPV model is assumed to be frozen-equivalent to the original model (or the real system) at frozen values of $p$; otherwise the difference between the output of $\Sigma$ and $\hat{\Sigma}$ does not approach zero even if the interval on which the scheduling signal is constant was very long. Instead, the difference cannot become smaller than a certain value, which is affected by the distance between the frozen models of $\Sigma$ and $\hat{\Sigma}$. Studying the effect of having an upper bound on the distance between the frozen models is left for future work. In this example, a linear interpolation is performed, as it is sufficient to give us the required model. For different systems, higher complexity models and interpolation methods might be required in order to avoid getting frozen-non-equivalent LPV models.

	\section{CONCLUSIONS}\label{sec:conclusion}
	\red{In this paper, an analytical expression for the difference between any two frozen-equivalent LPV models is presented. This difference could be seen as a modeling error of LPV systems, and the introduced expression allows us to evaluate the modeling error of LPV systems numerically, and gives the controller designer a bound on the global difference between frozen-equivalent LPV models. Future work will be aimed at using this expression in an optimization criteria during the identification process, more precisely, during the search for a coherent basis for the identified frozen models, which helps in minimizing the modeling error of LPV systems.}
	

%
	\bibliography{BIBIFAC}
	\appendix
	\section{Proofs}
	\subsection{Proof of Lemma~\ref{lem:continuousIsomorphism}}\label{prf:continuousIsomorphism}
     Note that for all $\bfp \in \Pset$, $T_{\hat{\Sigma},\Sigma}(\bfp)$ isomorphism satisfies
	\begin{align}\label{eq:TdefinitionReachability}
	T_{\hat{\Sigma},\Sigma}(\bfp) = &\mathcal{R}(\mathcal{L}_{\Sigma}(\bfp)) \mathcal{R}^\top(\mathcal{L}_{\hat{\Sigma}}(\bfp))\times  \\ & \hspace{50pt}[\mathcal{R}(\mathcal{L}_{\hat{\Sigma}}(\bfp)) \mathcal{R}^\top(\mathcal{L}_{\hat{\Sigma}}(\bfp))]^{-1}, \nonumber
	\end{align}
	where for all $\bfp \in \Pset$,  
	\begin{align}
         \mathcal{R}(\mathcal{L}_{\Sigma}(\bfp)) & \!\!=\!\! \begin{bmatrix}B(\bfp) & A(\bfp) B(\bfp) & \dots & A(\bfp)^{\NX-1} B(\bfp) \end{bmatrix},  \label{eq:TdefinitionReachabilitySigma} \\
         \mathcal{R}(\mathcal{L}_{\hat{\Sigma}}(\bfp)) &\!\!=\!\!  \begin{bmatrix}\hat{B}(\bfp) & \hat{A}(\bfp) \hat{B}(\bfp) & \dots & \hat{A}(\bfp)^{\NX-1} \hat{B}(\bfp) 
	\end{bmatrix}\label{eq:TdefinitionReachabilityHatSigma}
        \end{align}
	are the reachability matrices of $\mathcal{L}_{\Sigma}(\bfp)$ and $\mathcal{L}_{\hat{\Sigma}}(\bfp)$ respectively.
	Since $A, B, C$ and $\hat{A},\hat{B},\hat{C}$ are continuous maps, then from Eq.~\eqref{eq:TdefinitionReachabilitySigma} and Eq.~\eqref{eq:TdefinitionReachabilityHatSigma}, we can conclude that the maps $\Pset \ni \bfp \mapsto \mathcal{R}(\mathcal{L}_{\Sigma}(\bfp))$ and $\Pset \ni \bfp \mapsto \mathcal{R}(\mathcal{L}_{\hat{\Sigma}}(\bfp))$
are continuous. Hence, from Eq.~\eqref{eq:TdefinitionReachability}, $T_{\hat{\Sigma},\Sigma}$ is also a continuous map.
	
	\subsection{Proof of Lemma~\ref{lem:quadraticStability}}\label{prf:quadraticStability}
	\begin{pf}
		From the definition of quadratic stability of $\Sigma$, $\exists P$ positive definite matrix, such that the LMI
		\begin{align} \label{eq:qsLMI}
		A(\bfp)^{\top} P A(\bfp) - P < 0
		\end{align}
		holds for $\bfp \in \Pset$. Then, let us choose $S$ such that, $P = S^\top S$ and define $\hat{A}(\bfp):=S A(\bfp) S^{-1}$. Then, if follows directly that, $\forall \bfp \in \Pset$ 
		\begin{align}
		\APhat^\top \APhat - I_{\NX} &< 0,
		\end{align}
		\emph{i.e.},
		\begin{align}
		\Norm{\APhat} < 1.
		\end{align}
		Since $\Pset$ is compact, then, 
		\begin{align}
		\sup\limits_{\bfp \in \mathbb{P}}\Norm{\APhat} = \Norm{\hat{A}(\bfp_m)}
		\end{align} 
		for some $\bfp_m \in \Pset$ and the statement of the Lemma follows directly.
	\end{pf}
	\subsection{Proof of Theorem~\ref{th:differenceCharacterization}}\label{prf:th:differenceCharacterization}
	
	In order to simplify the equations, we define the following quantities, for each $\Delta >0, k\geq 0$ and $i \in \{0,1,\ldots,\Delta-1\}$:
	\begin{align*}
	\tks = k\Delta, & &
	\tke = k\Delta+\Delta-1, & &
	\tki = k\Delta+i.
	\end{align*}
	The intuition behind this notation is as follows. Consider any $p \in \mathcal{S}_{\Delta}$. Then $p$ is a piece-wise constant scheduling signal. Then $t_{k_s}$, $t_{k_e}$ and $t_{k_i}$ denote the starting time instance, the ending time instance and the $i^{th}$ time instance of the $k^{th}$ interval of the scheduling signal $p$ respectively. We illustrate these time instances in Figure~\ref{fig:timeAxis}. 
	
	\begin{figure}[ht]
		\centering
		\begin{picture}(190,50)(0,-7)
		\linethickness{0.3mm}
		\put(95,12){\line(1,0){95}}
		\linethickness{0.3mm}
		\put(0,12){\line(1,0){95}}
		\linethickness{0.3mm}
		\put(0,7){\line(0,1){7}}
		\linethickness{0.3mm}
		\put(190,7){\line(0,1){7}}
		\linethickness{0.3mm}
		\put(96,7){\line(0,1){7}}
		\linethickness{0.3mm}
		\put(93,7){\line(0,1){7}}
		\put(140,22){\makebox(0,0)[cc]{$k^{th}$ interval}}
		
		\put(48,22){\makebox(0,0)[cc]{$(k-1)^{th}$ interval}}
		
		\put(83,0){\makebox(0,0)[cc]{$\tkem$}}
		\put(105,0){\makebox(0,0)[cc]{$t_s^k$}}
		\put(145,0){\makebox(0,0)[cc]{$t_i^k$}}
		\put(189,0){\makebox(0,0)[cc]{$t_e^k$}}
		
		\put(8,0){\makebox(0,0)[cc]{$t_s^{k-1}$}}
		\put(40,0){\makebox(0,0)[cc]{\dots}}
		\put(120,0){\makebox(0,0)[cc]{\dots}}
		\put(160,0){\makebox(0,0)[cc]{\dots}}
		\put(131,33){\makebox(0,0)[cc]{}}
		
		\put(93,29){\makebox(0,0)[cc]{}}
		
		\end{picture}
		\caption{Scheduling signal time instances $\tks, \tke$ and $\tki$.}
		\label{fig:timeAxis}
	\end{figure}
	Note that on each set $\{\tks,\tks+1,\cdots, \tke\}$, the signal $p$ is constant, and $\tks = \tkem+1$, \emph{i.e.}, the sets $\{\tks,\tks+1,\cdots, \tke \}$ are adjacent and cover the whole time axis. 
	
		To derive the equations in the statement of the theorem, which characterize the differences between $\Sigma$ and $\hat{\Sigma}$, let $(x, y, p, u)$ be a solution of 
		\begin{align}\label{eq:sigma}
		\Sigma\left\{
		\begin{array}{lcl}
		x(t+1) &=&  A(p(t)) x(t) + B(p(t)) u(t) , \\
		y(t) &=& C(p(t)) x(t),
		\end{array}\right.
		\end{align}
		such that $x(0)=0$ and let $(\hat{x}, \hat{y}, {p}, {u})$, $\hat{x}(0)=0$ be a solution of 
		\begin{align}\label{eq:sigmaHat}
		\hat{\Sigma}\left\{
		\begin{array}{lcl}
		\hat{x}(t+1) &=& \hat{A}(p(t)) \hat{x}(t) + \hat{B}(p(t))u(t),\\
		\hat{y}(t) &=& \hat{C}(p(t)) \hat{x}(t).
		\end{array}\right.
		\end{align}
		
		In order to be able to compare the states of the two models $\Sigma$ and $\hat{\Sigma}$, we need to apply a coordinates transformation to transform the $\hat{\Sigma}$ to the same state basis of $\Sigma$. Define, $\forall i\in \{0,1,\cdots,\Delta-1\}$,
		\begin{align}\label{eq:xbarxhat}
		\bar{x}(\tki) = T_{\hat{\Sigma},\Sigma}(p(\tki))\hat{x}(\tki),
		\end{align}
		%
		where $T_{\hat{\Sigma},\Sigma}(p(\tki))$ is the isomorphism from the frozen model $\mathcal{L}_{\hat{\Sigma}}(p(\tki))$ to $\mathcal{L}_{\Sigma}(p(\tki))$.	Remember that $p(\tki)$ is constant on the set $\{\tks,\tks+1, \cdots, \tke\}$, \emph{i.e.}, $p(\tki) = p(\tks)$, then $\xbar{\tki} = \Tp{\tks}\xhat{\tki}$ for all $\tki$. 
		
		In order to characterize the difference between $y(\tki)={Y}_{\Sigma}(u,p)(\tki)$ and $\hat{y}(\tki)={Y}_{\hat{\Sigma}}(u,p)(\tki)$ by means of the model-related coefficients and the input, we break down the process to three steps for the sake of readability. In the first step, we relate the difference between $x$ and $\bar{x}$ inside the set $\{\tks,\tks+1,\ldots, \tke\}$, to the difference between $x$ and $\bar{x}$ at the beginning of this interval. 
				
		In the second step we relate the difference between $x$ and $\bar{x}$ at the beginning  $(\tks)$ of each interval, to the difference between $x$ and $\bar{x}$ at the end $(\tkem)$ of the preceding interval. 
		
		Finally, in the third step, we use the results of the preceding two steps in a recursive manner to relate the difference between $x$ and $\bar{x}$ at any time instant $t$ to the difference between $x$ and $\bar{x}$ at time instant $\Delta -1$.
		
%
		
		Next, we implement  these three steps mathematically.
		
		\subsubsection{1) Step one: relating to first value of the interval.}
		Now, we will characterize the difference between the states of the two LPV models over one interval of the scheduling signal, \emph{i.e.}, when the scheduling signal is constant. In this characterization, we expect to see the effect of the initial states, \emph{i.e.}, the difference at start time instant of the current interval of the scheduling signal, in addition to the length of the time interval $\Delta$.

		From the first equation in Eq.~\eqref{eq:sigmaHat} and the fact that
 		$\Ap{\tki}=T_{\hat{\Sigma},\Sigma}(p(\tki))\hat{A}(p(\tki))T^{-1}_{\hat{\Sigma},\Sigma}(p(\tki))$, 
		$\Bp{\tki}=T_{\hat{\Sigma},\Sigma}(p(\tki))\hat{B}(p(\tki))$, it follows that, $\forall i\in \{0,1,\ldots \Delta-1\}$,
		\begin{align}
		\xbar{\tki+1}=\Ap{\tki}\xbar{\tki}+\Bp{\tki}u(\tki).
		\end{align}
		By subtracting above equation from the first equation in Eq.~\eqref{eq:sigma}
		we get,
		\begin{align}
		x(\tki+1)-\xbar{\tki+1} = \Ap{\tki}(x(\tki)-\xbar{\tki}).
		\end{align}
		Then, recursively, on the same time interval, we get 
		\begin{align}\label{eq:xminusxbar}
		x(\tki)-\xbar{\tki} = \Ap{\tks}^{i}(x(\tks)-\xbar{\tks}).
		\end{align}
		Next, by replacing Eq.~\eqref{eq:alphaA} in Eq.~\eqref{eq:xminusxbar}, we find that,
		\begin{align}\label{eq:xSameInterval}
		\Norm{x(\tki)-\xbar{\tki}} & \leq \alpha^{i}\Norm{x(\tks)-\xbar{\tks}}.
		\end{align}\unboldmath
		
		
		\subsubsection{2) Step two: relating to the precedent interval.}
		
		Now, in order to get a step backward on the time axis, we need to relate the initial values of the time interval with the last value of the previous time interval. For this purpose, define
		\begin{align*}\beta =& \Ap{\tkem}\xbar{\tkem}+\Bp{\tkem}u(\tkem).\end{align*}
		Notice that $\beta$ is a constant value, and one should not confuse it with a state trajectory.
		

		By taking the norm of $\beta$, we get
		\begin{align*}
		\Norm{\beta} & \leq \alpha K_T \mu_1 \Normind{u}{l_{\infty}} + K_B\Norm{u(\tkem)}.
		\end{align*}
		Notice, from the definition of the $l_{\infty}$ norm, 
		\begin{align}
		\forall k \in  \mathbb{N}: & \,\,\,\, \Normind{u}{l_{\infty}} \geq \Norm{u(k)} .
		\end{align}
		Then,
		\begin{align}\label{eq:kbeta}
		\Norm{\beta} & \leq  (\alpha K_T \mu_1 + K_B).\Normind{u}{l_{\infty}} .
		\end{align} 
		Now, using the triangle inequality, we can write
		\begin{align}
		\Norm{x(\tks)-\xbar{\tks}} \leq \Norm{x(\tks)-\beta}+\Norm{\beta-\xbar{\tks}}.
		\end{align}
		Notice that,
		\begin{align*}
		\Norm{x(\tks)-\beta} &=\Norm{\Ap{\tkem}(x({\tkem})-\xbar{\tkem})}\\
		& \leq \Norm{\Ap{\tkem}}. \Norm{x(\tkem)-\xbar{\tkem}}.
		\end{align*}
		Then,
		\begin{align}\label{eq:xbetapart1}
		{\Norm{x(\tks)-\beta} \leq \alpha \Norm{x(\tkem)-\xbar{\tkem}}}.
		\end{align}
		Remember that
		\begin{align}
		\xbar{\tks} =& \Tp{\tks}\xhat{\tks} \nonumber\\
		=& \Tp{\tks}[\Aphat{\tkem}\xhat{\tkem}+\Bphat{\tkem}u(\tkem)]\nonumber\\
		=& \Tp{\tks}\Tpinv{\tkem}[\Ap{\tkem}\xbar{\tkem}\nonumber\\&+\Bp{\tkem}u(\tkem)],
		\end{align}
		\emph{i.e.},
		\begin{align}
		\xbar{\tks} =& M_{p(\tks),p(\tkem)}\beta .
		\end{align}
		Along with Eq.~\eqref{eq:xbarxhat}, we can directly find that 
		\begin{align}
		\beta-\xbar{\tks} =& \Big(I-M_{p(\tks),p(\tkem)}\Big)\beta .
		\end{align}
		Then,
		\begin{align}\label{eq:betaXbarbeta}
		\Norm{\beta - \xbar{\tks}} &\leq \Norm{I-M_{p(\tks), p(\tkem)}}.\Norm{\beta}.	
		\end{align}
		Then, by replacing \eqref{eq:kbeta} and \eqref{eq:km} in Eq.~\eqref{eq:betaXbarbeta},we get
		\begin{align}\label{eq:xbetapart2}
		{\Norm{\beta - \xbar{\tks}}	\leq K_M(p)(\alpha K_T \mu_1 + K_B) \Normind{u}{l_{\infty}}.}
		\end{align}
		Then, combining \eqref{eq:xbetapart1} and \eqref{eq:xbetapart2}, we find that
		\begin{align}\label{eq:xPreviousInterval}
		\Norm{x(\tks)-\xbar{\tks}} \leq & \alpha \Norm{x(\tkem)-\xbar{\tkem}} \\ &+ K_M(p)(\alpha K_T \mu_1 + K_B) \Normind{u}{l_{\infty}} \nonumber.
		\end{align}\unboldmath
		
		In the last step, we recursively find the relation of the output difference with the  initial state of the systems.
		\subsubsection{3) Step three: recursivity.}
		From Eq.~\eqref{eq:xSameInterval} and Eq.~\eqref{eq:xPreviousInterval}, we can  recursively find that
		\begin{align}
		\Norm{x(\tki)-\xbar{\tki}} \leq \,\, &\alpha^{i} \alpha^{(k-1)\Delta}\Norm{x(\Delta-1)-\xbar{\Delta-1}} \nonumber \\&+\alpha^{i}\sum_{j=0}^{k-1}\alpha^{\Delta j}K_u(p) \Normind{u}{l_{\infty}},
		\end{align}
		where $K_u(p) = K_M(p)(\alpha K_T \mu_1 + K_B)$. Notice that $T_{\hat{\Sigma},\Sigma}(p(0))$ is an isomorphism from $\mathcal{L}_{\hat{\Sigma}}(p(0))$ to $\mathcal{L}_{\Sigma}(p(0))$, \emph{i.e.}, the local models of $\hat{\Sigma}$ and $\Sigma$ during the first $\Delta$ time instances. 
This yields directly that, 
		\[\Norm{x(\Delta-1)-\xbar{\Delta-1}} = 0.\] 
		Then,
		\begin{align}\label{eq:xRes}
		\Norm{x(\tki)-\xbar{\tki}} \leq \,\, & \alpha^{i}K_u(p) \Normind{u}{l_{\infty}}\sum_{j=0}^{k-1}\alpha^{\Delta j}.
		\end{align}
		Notice that, for $\alpha \in [0,1]$, we have
		\begin{align}
		{\sum_{i=0}^{k-1}\alpha^{\Delta j} \leq \sum_{j=0}^{\infty}\alpha^{\Delta j}=\frac{1}{1-\alpha^{\Delta}}}.
		\end{align}
		Then, replacing in the equation \eqref{eq:xRes}, we get
		\begin{align}\label{eq:xbary}
		{\Norm{x(\tki)-\xbar{\tki}}  \leq \frac{\alpha^{i}}{1-\alpha^{\Delta}}K_u(p) \Normind{u}{l_{\infty}}}.
		\end{align}\unboldmath
		Then,
		notice that, using Eq~\eqref{eq:xbarxhat}, we can write
		\begin{align}
			C(p(\tki))\xbar{\tki} &= C(p(\tki))T_{\hat{\Sigma},\Sigma}(p(\tki))\hat{x}(\tki) \\&= \Chat{\tki}\hat{x}(\tki)  = \hat{y}(\tki). \nonumber
		\end{align}
		Then, from Eq. ~\eqref{eq:xbary}, after multiplication by $C(p(\tki))$, we get that,
		\begin{align*}
		\Norm{y(\tki)\! -\!\hat{y}(\tki)}
		\!\! \leq\!\! \frac{\alpha^{i}}{1-\alpha^{\Delta}}K_u(p) \sup\limits_{p} \Norm{C(p)}. \Normind{u}{l_{\infty}} .
		\end{align*}
		\emph{i.e.},
		\begin{align*}
		\Norm{Y_{\Sigma}(u, p)(\tki) -Y_{\hat{\Sigma}}(u,p)(\tki)} \leq \gfunction{\Delta}{K_M(p)}{i}. \Normind{u}{l_{\infty}}.
		\end{align*}\unboldmath
		Remember that for all $p \in \mathcal{P}$, $K_M(p) \leq K_M$, then,
		\begin{align*}
		\Norm{Y_{\Sigma}(u, p)(\tki) -Y_{\hat{\Sigma}}(u,p)(\tki)} \leq \gfunction{\Delta}{K_M}{i}. \Normind{u}{l_{\infty}}.
		\end{align*}
	\subsection{Proof of Theorem~\ref{th:switchingInterval}}\label{prf:th:switchingInterval}
	
		Remember that $0<\alpha < 1$, then,
		\begin{align}
		\lim\limits_{\boldsymbol{i \rightarrow \infty}}\frac{\alpha^{i}}{1-\alpha^{\Delta}} = 0.
		\end{align}
		Then, from Eq.~\eqref{eq:gDefinition}, we can find that 
		\begin{align}\lim\limits_{\boldsymbol{i \rightarrow \infty}}\gfunction{\Delta}{K_M(p)}{i} = 0.\end{align}
		This means, $\forall \varepsilon> 0, \exists \Delta_m > 0$, such that,
		\begin{align}
		\forall i\geq \Delta_m: \gfunction{\Delta}{K_M(p)}{i} < \varepsilon,
		\end{align}
		Then, from Eq~\eqref{eq:outputEquation}, $\forall i > \Delta_m$: 
		\begin{align*}
		\Norm{Y_{\Sigma}(u, p)(k\Delta+i) -Y_{\hat{\Sigma}}(u,p)(k\Delta+i)}\\
	& \hspace{-50pt}\leq \gfunction{\Delta_m}{K_M(p)}{i} . \Normind{u}{l_{\infty}} \\& \hspace{-50pt} \leq \varepsilon . \Normind{u}{l_{\infty}}.
		\end{align*}
	
	\subsection{Proof of Theorem~\ref{th:speedOfChange}}\label{prf:th:speedOfChange}
		Consider the map
		\begin{align}
		\varphi: \Pset \times \Pset \mapsto \mathbb{R}^{\NX \times \NX}: (\bfp_1
		,\bfp_2) \rightarrow I-M_{\bfp_1,\bfp_2}.
		\end{align}
		Since $T_{\hat{\Sigma},\Sigma}$ is continuous, so is $T_{\hat{\Sigma},\Sigma}^{-1}$ and hence $\varphi$ is continuous.	Remember that  $\Pset$ is compact, then so is $\Pset \times \Pset$. As $\varphi$ is a continuous function defined on a compact set, then it is uniformly continuous.
		
		In particular, since $\varphi(\bfp_1, \bfp_1) = 0$ for all $\bfp_1 \in \Pset$, $\forall \varepsilon_1 > 0: \exists \delta(\varepsilon_1) > 0$, such that, if $\Norm{\bfp_1 - \bfp_2} = \Norm{(\bfp_1 - \bfp_2)-(\bfp_1 - \bfp_1)} < \delta(\varepsilon_1)$, then,
		\begin{align}
			\Norm{\varphi(\bfp_1, \bfp_2) - \varphi(\bfp_1, \bfp_1)} = \Norm{\varphi(\bfp_1, \bfp_2)} < \varepsilon_1.
		\end{align}
		Hence, if $p \in \mathcal{P}$ is such that $\Norm{p(t+1) - p(t)} < \delta(\varepsilon_1)$, then $K_M(p) < \varepsilon_1$.

		Then, we can take
		\begin{align}
			\varepsilon_1 = \varepsilon \Big(\frac{\alpha}{1-\alpha^{\Delta}}(\alpha K_T \mu_1 + K_B)K_C\Big)^{-1}.
		\end{align}
		This means that,
		\begin{align*}
			\gfunction{\Delta}{K_M(p)}{1} =& \frac{\alpha}{1-\alpha^{\Delta}}K_M(p) (\alpha K_T \mu_1 + K_B) K_C \\
			&\leq \varepsilon_1 \frac{\alpha}{1-\alpha^{\Delta}} (\alpha K_T \mu_1 + K_B) K_C
			\\
			& = \varepsilon.
		\end{align*}
		Then, from Eq~\eqref{eq:outputEquation}, with $\Delta = 1$, $\forall \varepsilon > 0,  \exists \delta>0$ such that  if $\Norm{p(t+1)-p(t)} < \delta$, then,
		\begin{align*}
		\Norm{Y_{\Sigma}(u, p)(t) -Y_{\hat{\Sigma}}(u,p)(t)}&\leq \gfunction{\Delta}{K_M(p)}{1}.\Normind{u}{l_{\infty}}  \\&\leq \varepsilon . \Normind{u}{l_{\infty}}.
		\end{align*}

	\subsection{Proof of Theorem~\ref{th:stability}}\label{prf:th:stablility}
		Notice that, from Eq.~\eqref{eq:gDefinition}, we can find that,
		\begin{align}\lim\limits_{\boldsymbol{\alpha \rightarrow 0}}\gfunction{\Delta}{K_M(p)}{i} = 0.\end{align}
		This means that $\forall \varepsilon > 0, \exists \alpha_m >0$, such that, $\forall \alpha < \alpha_m$,
		\begin{align}
		\gfunction{\Delta}{K_M(p)}{i} < \varepsilon.
		\end{align}
		Then, from Eq.~\eqref{eq:outputEquation}, in additions to our assumption that $\Norm{A}< \alpha$, we directly get that, if $\alpha < \alpha_m$, then, $ \forall t>0$,
		\begin{align*}
		\Norm{Y_{\Sigma}(u, p)(t) -Y_{\hat{\Sigma}}(u,p)(t)}& \leq \gfunction{\Delta}{K_M(p)}{i} . \Normind{u}{l_{\infty}} \\ &\leq \varepsilon. \Normind{u}{l_{\infty}}.
		\end{align*}
\end{document}